\documentclass[twocolumn,showpacs,showkeys,preprintnumbers,amsmath,amssymb]{revtex4}
 \usepackage{dcolumn}
 \usepackage{bm}
\usepackage{graphicx}
\usepackage{subfigure}
 \begin{document}

 \title{Scalar clouds in charged
  stringy black hole-mirror system}

 \author{Ran Li}

 \thanks{Corresponding author. Electronic mail: 021149@htu.cn}

 \author{Junkun Zhao}

 \author{Xinghua Wu}

 \author{Yanming Zhang}

 \affiliation{Department of Physics,
 Henan Normal University, Xinxiang 453007, China}

 \begin{abstract}

 It was reported that massive scalar fields can form bound states around Kerr black holes
 [C. Herdeiro, and E. Radu, Phys. Rev. Lett. 112, 221101 (2014)].
 These bound states are called scalar clouds, which have a real
 frequency $\omega=m\Omega_H$, where $m$ is the azimuthal index
 and $\Omega_H$ is the horizon angular velocity of Kerr black hole.
 In this paper, we study scalar clouds in a spherically symmetric background,
 i.e. charged stringy black holes,
 with the mirror-like boundary condition. These bound states satisfy
 the superradiant critical frequency condition $\omega=q\Phi_H$
 for charged scalar field, where $q$ is charge
 of scalar field, and $\Phi_H$ is horizon electrostatic potential.
 We show that, for the specific set of
 black hole and scalar field parameters, the clouds are only possible for
 the specific mirror locations $r_m$. It is shown that
 analytical results of mirror location $r_m$ for the clouds
 are perfectly coincide with numerical results in the $qQ\ll 1$ regime.
 We also show that the scalar clouds are also possible when the mirror locations are
 close to the horizon. At last, we provide an analytical calculation 
 of the specific mirror locations $r_m$ for the scalar clouds in the $qQ\gg 1$ regime.

 \end{abstract}

 \pacs{04.70.-s, 04.60.Cf}

 \keywords{black hole, superradiance, scalar cloud}

 \maketitle

 \section{Introduction}

 It was firstly proposed by S. Hod that the scalar field can have real
 bound states in the near-extremal Kerr black hole \cite{hodprd2012,hodepjc2013}. Soon later,
 it was reported in \cite{herdeiroprl} that massive scalar fields can form bound states around Kerr black holes by using the numerical method to solve the scalar field equation in the background.
 This bound states are the stationary scalar configurations in the black hole backgrounds,
 which are regular at the horizon and outside. They are named as
 scalar clouds. More importantly, it was shown that the backreaction
 of clouds can generate a new family of Kerr black holes with scalar hair \cite{herdeiroprl,herdeiroprd}.
 It is suggested that whenever clouds of a given matter
 field can be found around a black hole, in a linear analysis, there
 exists a fully non-linear solution of new hairy black hole correspondingly.
 However, it requires that the field originating clouds yields a time 
 independent energy momentum tensor. Generally, the field should be complex, and 
 have a factor $e^{-i\omega_c t}$, where $\omega_c$ is the superradiance critical 
 frequency. For instance, real scalar fields
 can give rise to clouds but not hairy black holes \cite{1501}.
 So, it seems that the studies of scalar clouds in the linear level
 are very important for us to find the
 hairy black holes in the non-linear level.
 This subject has attracted a lot of attention recently
 \cite{hodprd2014,RNclouds,benone,sampaio,grahamprd,Degolladoclouds,
 Brihaye,HRRPLB,hodplb1,hodplb2,acoustic}.

 Generally speaking, the existence of stationary bound states
 of matter fields in the black hole backgrounds requires two
 necessary conditions.
 The first is that the matter fields should undergo the classical
 superradiant phenomenon \cite{bardeen,misner} in the black hole background. This
 condition can be satisfied by the bosonic fields in the rotating
 black holes or the charged scalar fields in the charged
 black holes \cite{bekenstein}. When the frequencies of these matter fields $\omega$
 are smaller than the superradiant critical frequency $\omega_c$,
 there are time growing quasi-bound states. When $\omega>\omega_c$,
 the fields are time decaying. So, the scalar clouds exists at the
 boundary between these two regimes, i.e. the frequencies of the fields
 are taken as the superradiant critical frequency $\omega_c$.
 For the rotating black holes,
 the critical frequency $\omega_c$ is
 $m\Omega_H$, where $m$ is the azimuthal index
 and $\Omega_H$ is the horizon angular velocity.
 While for the charged black holes, $\omega=q\Phi_H$,
 where $q$ is the charge
 of scalar field, and $\Phi_H$ is the horizon electrostatic potential.
 The second one is there should be a potential well
 outside the black hole horizon in which the bound states can be trapped.
 This potential well may be
 provided by the mass term of the field, i.e. $\omega<\mu$,
 where $\mu$ is the mass of the scalar field. However, sometimes
 the artificial boundary conditions can also play the same role.

 In this paper, we will study the scalar clouds in a spherically symmetric
 and charged background. Specifically, we will consider the charged scalar field
 in the backgrounds of the charged stringy black holes. At first sight,
 it seems that the massive scalar field can form the clouds in this background.
 However, it is proved that the the massive charged scalar field
 is stable in this background and there is no superradiant instability \cite{liprd}.
 To generate the superradiant instability \cite{detweiler}, the mirror-like boundary condition
 should be imposed according to the black hole bomb mechanism \cite{press,cardoso2004bomb}.
 The analytical and the numerical studies on this subject can be found in
 \cite{liepjc2014} and \cite{liplb2015}.
 Correspondingly, the scalar clouds are only possible
 with the mirror-like boundary condition. Using the numerical method,
 we will study the dynamics of the massless charged scalar field satisfying
 the frequency condition $\omega=q\Phi_H$ and the mirror-like boundary condition.
 We will show that, for the specific set of
 black hole and scalar field parameters, the clouds are only possible for
 the specific mirror locations $r_m$. It will be shown that
 the analytical results of mirror location $r_m$ for the clouds
 are perfectly coincide with the numerical results. In addition, we will show that
 the scalar clouds are also possible when the mirror locations are
 close to the horizon. At last, we will provide an analytical calculation
 of the specific mirror locations $r_m$ for the scalar clouds in the $qQ\gg 1$ regime.

 This paper is organized as follows. In Sec.II,
 we will present the background geometry of charged string black hole
 and the dynamic equation of the scalar field. In particularly,
 we will give the superradiant condition and the boundary
 condition of this black hole-mirror system.
 In Sec.III, we describe the numerical procedure to solve the
 radial equation under the certain boundary condition.
 In this section, the numerical results are also
 illustrated. Some general discussion on the numerical
 results are also followed. In Sec. IV, an analytical calculation
 of the mirror radius $r_m$ for scalar clouds in $qQ\gg 1$ regime
 is present. The conclusion is appeared in Sec. IV.

 \section{Description of the system}

 We shall consider a massless charged scalar field minimally coupled to
 the charged stringy black hole with the mirror-like boundary condition.
 The black hole is a static spherical symmetric charged black holes
 in low energy effective theory
 of heterotic string theory in four dimensions, which is firstly found by
 Gibbons and Maeda in \cite{GM} and independently
 found by Garfinkle, Horowitz, and Strominger in \cite{GHS} a few years later.
 The metric is given by
 \begin{eqnarray}
 ds^2&=&-\left(1-\frac{2M}{r}\right)dt^2+\left(1-\frac{2M}{r}\right)^{-1}
 dr^2\nonumber\\
 &&+r\left(r-\frac{Q^2}{M}\right)(d\theta^2+\sin^2\theta d\phi^2)\;,
 \end{eqnarray}
 and the electric potential and the dilaton field
 \begin{eqnarray}
 &&A_t=-\frac{Q}{r}\;,\;\;\\
 &&e^{2\Phi}=1-\frac{Q^2}{Mr}\;.
 \end{eqnarray}
 The parameters $M$ and $Q$ are the mass and the electric charge
 of the charged stringy black hole, respectively.
 The event horizon of black hole is located at $r=2M$.
 The area of the sphere approaches to zero when $r=Q^2/M$. Therefore, the
 sphere surface of the radius $r=Q^2/M$ is singular. When $Q^2\leq 2M^2$,
 this singular surface is surrounded by the event horizon.
 In this paper, we will always assume
 the cosmic censorship hypothesis, i.e.
 we will only consider the black hole with the parameters
 satisfying the condition $Q^2\leq 2M^2$.

 The dynamics of the charged scalar field
 is then governed by the Klein-Gordon equation
 \begin{eqnarray}
 (\nabla_\nu-iqA_\nu)(\nabla^\nu-iqA^\nu)\Psi=0\;,
 \end{eqnarray}
 where $q$ denotes the charge of the scalar field.
 By taking the ansatz of the scalar field
 $\Psi=e^{-i\omega t}R(r)Y_{lm}(\theta,\phi)$,
 where $\omega$ is the conserved energy of the mode,
 $l$ is the spherical harmonic index, and $m$ is the
 azimuthal harmonic index with $-l\leq m\leq l$,
 one can deduce the radial wave equation in the form of
 \begin{eqnarray}
 \Delta\frac{d}{dr}\left(\Delta \frac{dR}{dr}\right)+UR=0\;,
 \end{eqnarray}
 where we have introduced a new function $\Delta=\left(r-r_+\right)\left(r-r_-\right)$
 with $r_+=2M$ and $r_-=Q^2/M$,
 and the potential function is given by
 \begin{eqnarray}
 U=\left(r-\frac{Q^2}{M}\right)^2(\omega r-qQ)^2-\Delta l(l+1)\;.
 \end{eqnarray}

 The superradiant condition of the charged scalar field is given by
 \begin{eqnarray}
 \omega<q\Phi_H\;,
 \end{eqnarray}
 where $\Phi_H=\frac{Q}{2M}$ is the electric potential at the
 horizon\cite{dilatonsr,liprd}. It is proved in \cite{liprd} that the massive charged
 scalar field is stable in this black hole background. To have superradiant
 instability, we should impose the mirror-like boundary condition \cite{liepjc2014, liplb2015}.
 In order to study the bound states, we shall focus on the critical case that
 the scalar frequency equals to the superradiant critical frequency, i.e.
 \begin{eqnarray}
 \omega=q\Phi_H\;.
 \end{eqnarray}

 To solve the radial equation (5), we should impose the following boundary conditions,
 which are given by
 \begin{eqnarray}
 R(r)=\left\{
        \begin{array}{ll}
          R_0\left(1+\sum_{k\geq 1}R_k(r-r_+)^k\right), & r\rightarrow r_+ \\
          0, & r=r_m
        \end{array}
      \right.
 \end{eqnarray}
 The first line indicates that the scalar field is regular near the horizon and the second line
 implies that the system is placed in a perfectly reflecting cavity.

 \section{Numerical procedure and results}

 The numerical methods employed in this problem are based on the shooting method,
 which is also called the direct integration (DI) method \cite{Degolladoprd, Dolanprd2010,
 Cardosoprd2014,Uchikata}. It is shown that the DI method is specially suited to find
 stationary field configuration with the mirror-like boundary condition.

 Firstly, near the event horizon $r=2M$, we require the radial function is regular
 and expand the radial function $R(r)$ as
 a generalized power series in terms of $(r-r_+)$
 as have done in the first line of Eq.(9). Because the radial equation
 is linear, we can take $R_0=1$ without loss of generality. Substituting expansion of the radial
 wave function into the radial equation (5), we can solve the coefficient $R_k$ order by order
 in terms of $(r-r_+)$. We have only considered six terms in the expansion. The $R_k$s
 can be expressed in terms of the parameters $(M, Q, q, l)$, which are not exhibited here.

 Then, we can integrate the radial equation (5) from  $r=r_+(1+\epsilon)$
 and stop the integration at the radius of the mirror. In this procedure,
 we have taken the small $\epsilon$ as $10^{-6}$. The procedure can be repeated
 by varying the input parameters $(M, Q, q, l)$ until the mirror-like boundary condition $R(r_m)=0$ is reached with the desired precision.
 We can use a numerical root finder to search the
 location of the mirror that support the stationary scalar configuration.
 We have found that, for the given input parameters $(M, Q, q, l)$,
 scalar clouds exist for a discrete set of $r_m$, which is labeled by the quantum number
 $n$ of nodes of the radial function $R(r)$.

\begin{figure}
\subfigure{\includegraphics{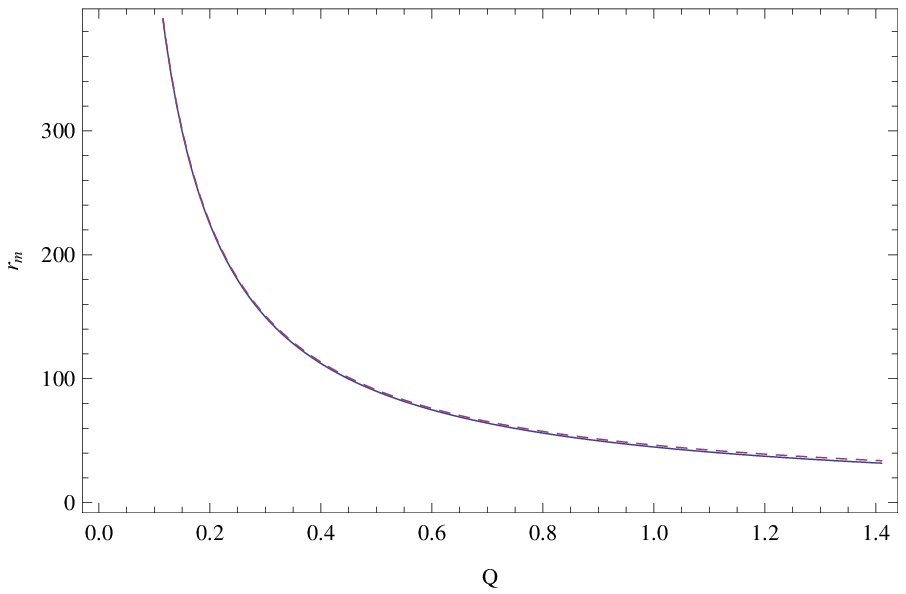}}
\subfigure{\includegraphics{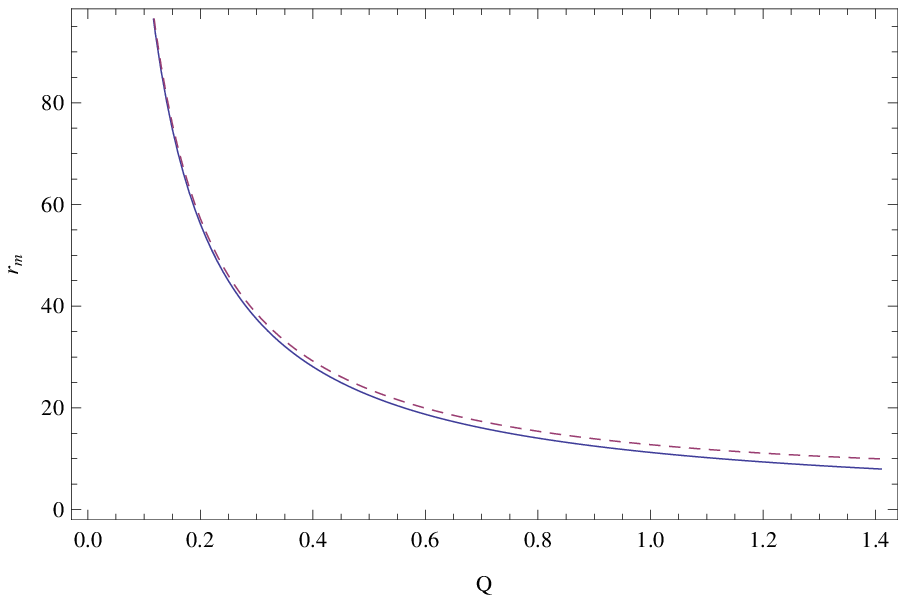}}
\caption{Mirror location $r_m$
plotted versus the black hole charge $Q$ for $M=1, l=1, n=0$ and for various scalar charge $q$.
For the first panel, $q=0.2$, while for the second panel, $q=0.8$.
The solid line and the dashed line represent the analytical and the numerical
results respectively.}
\end{figure}

 Firstly, we make a comparison of the numerical and analytical
 results. From the analytical result Eq.(35) in Ref.\cite{liepjc2014},
 one can obtain the mirror radius that supports scalar cloud can
 be approximately given by
 \begin{eqnarray}
 r_m=\frac{j_{l+1/2,n'}}{q\Phi_H}\;.
 \end{eqnarray}
 We have labeled the $n'$-th positive zero of the Bessel function $J_{l+1/2}$
 as $j_{l+1/2, n'}$. The numerical results show that this "quantum number"
 is closely connected with the nodes number $n$ of the radial function $R(r)$
 towards the simple relation $n=n'-1$.
 It should be noted that this analytical expression for the mirror radius
 is only valid for the case of $qQ\ll 1$. With the condition $qQ\ll 1$,
 the asymptotic expansion matched method can be employed to solve the radial equation
 approximately \cite{liepjc2014}.
 In Fig.(1), we have displayed the analytical results and the numerical results
 of the mirror location $r_m$ in terms of the black hole charge $Q$.
 Here, we do not consider the naked singularity spacetime, so that the
 value range of black hole charge $Q$ is $(0, \sqrt{2}]$,
 where  we have fixed the black hole mass as $M=1$.
 It is shown that the analytical results of mirror location $r_m$ for the clouds
 are perfectly coincide with the numerical results, even in the region
 where the analytical approximation is unapplicable.
 When $q=0.2$, the analytical approximation is always precise
 in all range of $Q$. When $q=0.8$, the analytical results have obvious
 difference with the numerical results only for large $Q$.

\begin{figure}
\subfigure{\includegraphics{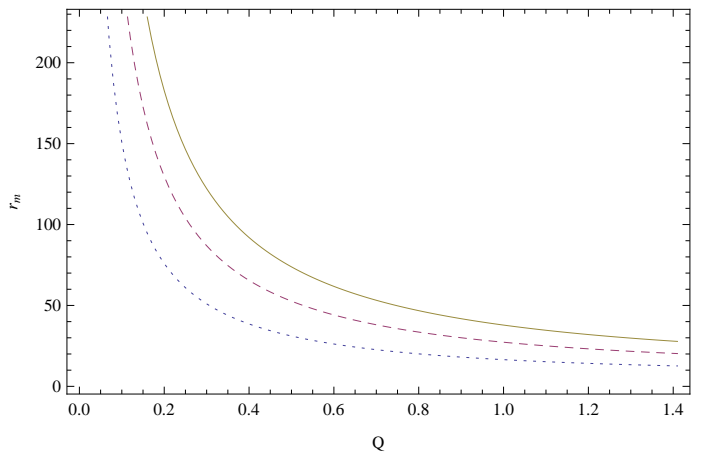}}
\caption{Mirror location $r_m$
plotted versus the black hole charge $Q$ for $M=1, l=1, q=0.6$ and for various node number $n$.
The dotted, dashed, and solid lines represent $n=0, 1$, and $2$ respectively.}
\end{figure}

 In Fig.(2), we have drawn the mirror location $r_m$ that support the scalar cloud
 as a function of the black hole charge $Q$ for various values of node number $n$ of the radial
 function. It is observed that, when the black hole charge $Q$ increases, we need to
 place the reflecting mirror more closer to the horizon in order to
 have a scalar cloud. When the node number $n$ of radial function increases,
 the plotted lines become away from the axis. This observation is coincide with the
 analytical result (10) in the regime of $qQ\ll 1$.

\begin{figure}
\subfigure{\includegraphics{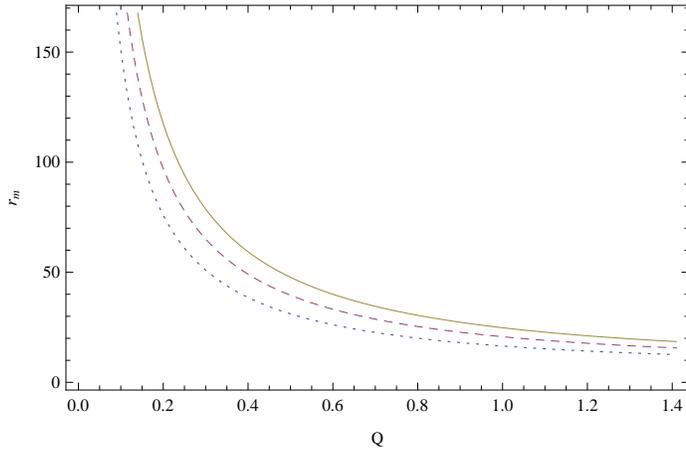}}
\caption{Mirror location $r_m$
plotted versus the black hole charge $Q$ for $M=1, n=0, q=0.6$ and for various $l$.
The dotted, dashed, and solid lines represent $l=1, 2$, and $3$ respectively.}
\end{figure}

 In Fig.(3) and (4), we display the mirror location $r_m$ as a function of
 the black hole charge $Q$ for various different $l$ and $q$.
 We can observe that, the lines become far away from the axis when increasing $l$,
 while the lines become more closer to the axis when increasing the scalar charge $q$.
 This is also expected from the analytical result (10). In addition,
 Fig.(3) and (4) together with Fig.(2) show that,
 when $Q\rightarrow 0$, $r_m\rightarrow \infty$. This indicates that
 there is no massless scalar cloud for Schwarzschild black hole
 with the mirror-like boundary condition \cite{acoustic},
 even thought it is possible for massive scalar fields in Schwarzschild
 black hole to have arbitrarily long-lived quasi-bound states \cite{barranco}.

\begin{figure}
\subfigure{\includegraphics{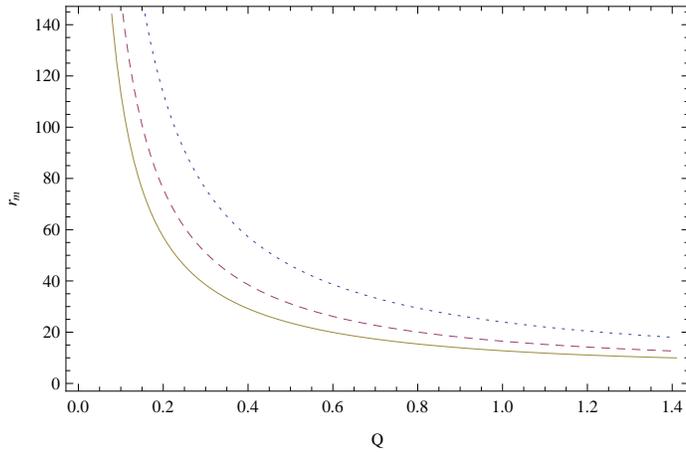}}
\caption{Mirror location $r_m$
plotted versus the black hole charge $Q$ for $M=1, l=1, n=0$ and for various scalar charge $q$. The dotted, dashed, and solid lines represent $q=0.4, 0.6$, and $0.8$ respectively.}
\end{figure}

\begin{figure}
\subfigure{\includegraphics{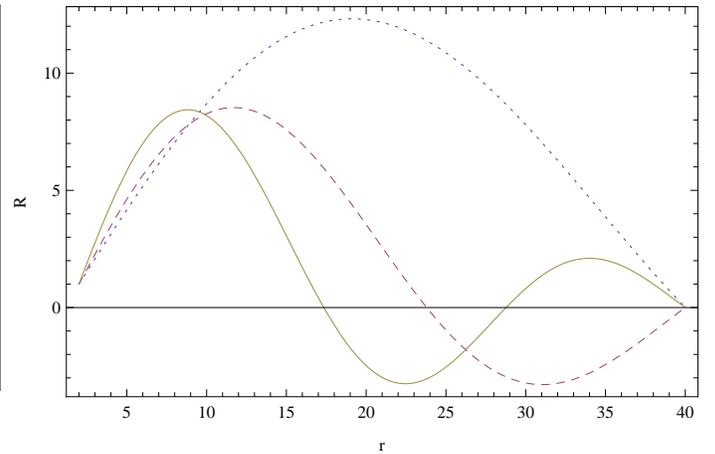}}
\subfigure{\includegraphics{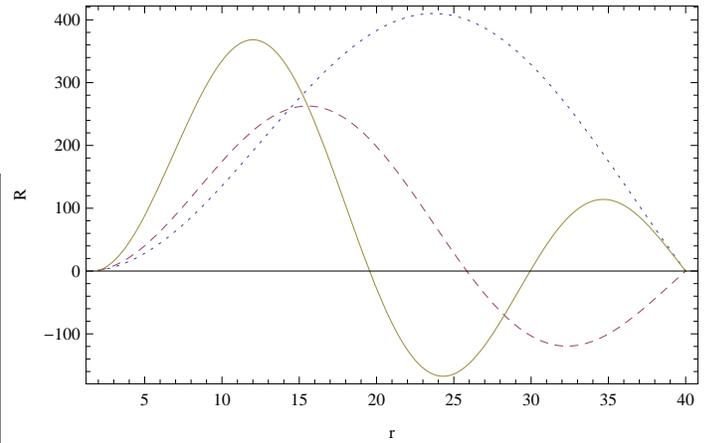}}
\caption{Radial functions $R(r)$ of scalar clouds
for $M=1, q=0.6, r_m=40$ with different harmonic index $l$ and
node number $n$. The first and the second panels correspond
$l=1$ and $2$ respectively. The dotted, dashed, and solid
lines represent $n=1, 2$, and $3$ respectively.}
\end{figure}

\begin{figure}
\subfigure{\includegraphics{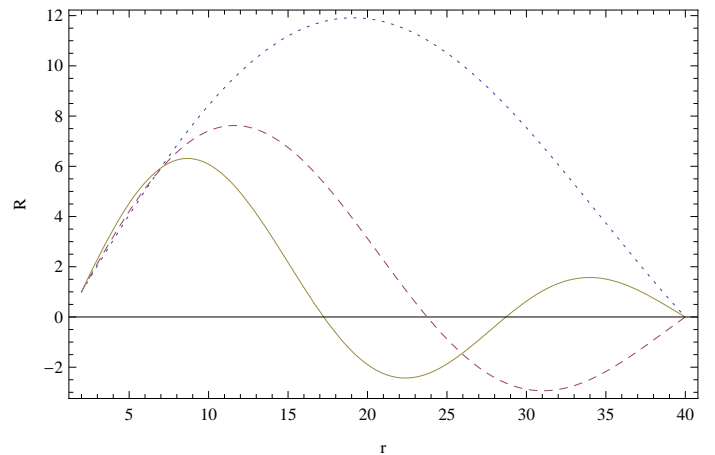}}
\caption{Radial functions $R(r)$ of scalar clouds
for $M=1, q=0.8, l=1, r_m=40$ with different
node number $n$. The dotted, dashed, and solid
lines represent $n=1, 2$, and $3$, respectively,
and the corresponding black hole charge $Q$ are
$0.219882, 0.583819$, and $0.956562$.}
\end{figure}

 We also consider the radial dependence of the massless
 scalar clouds. In Fig.(5) and (6),
 we have fixed the mirror radius as $r_m=40$.
 We can solve the radial equation numerically and
 obtain a discrete set of black hole charge $Q$ which is labeled
 by the node number $n$ of the radial wave equation.
 Then we can integrate the radial equation for the fixed node numbers
 and obtain the corresponding numerical solutions of the radial wave functions.
 It is shown that the radial profile
 have the typical forms of standing waves with the
 fixed boundary conditions. We have also calculated the case that $l=3$. 
 The results is not present here. The general form of the radial wave 
 function is similar to the profiles in Fig.(5).

 In Fig.(7), we consider the case that the mirror location
 is very close to the horizon. We take the mirror radius as $r_m=3$.
 From our previous analytical and numerical work
 on the superradiant instability of scalar field in the background
 of the charged stringy black hole plus mirror system,
 we need a large scalar field charge $q$. Here, we set $q=20$.
 We can see that, the scalar field can be bounded by the reflecting mirror
 very near the horizon to form the clouds. The radial wave function
 in this case have the similar profiles as Fig.(5) and (6).

\begin{figure}
\subfigure{\includegraphics{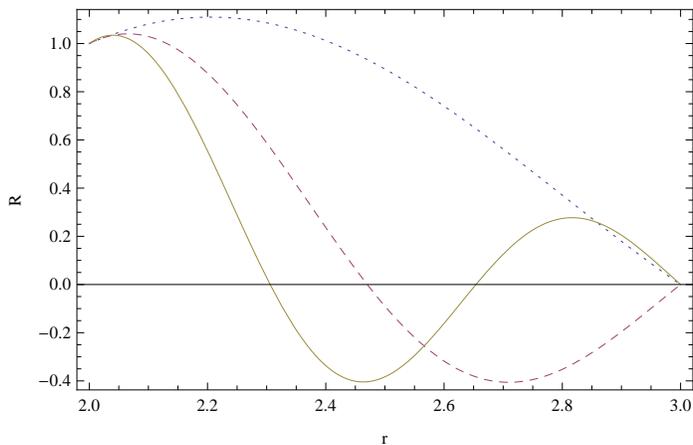}}
\caption{Radial functions $R(r)$ of scalar clouds
for the small mirror radius $r_m=3$. The parameters of
black hole and scalar field are taken as $M=1, q=20$, and $l=1$.
The dotted, dashed, and solid
lines represent $n=1, 2$, and $3$, respectively, and the
corresponding black hole charge $Q$ are $0.306384, 0.600699$,
$0.913741$.}
\end{figure}

 \section{Scalar clouds in $qQ\gg 1$ regime}

 In the above numerical calculations, we find the radial equation 
 becomes hard to integrate when the scalar charge $q$ is large. 
 So it is important to make an analytical study of the stationary 
 charged scalar clouds in the $qQ\gg 1$ regime. In this section, 
 we will give an analytical expression of special mirror radius $r_m$ 
 in $qQ\gg 1$ limit, for which the charged scalar field can be confined 
 to form stationary cloud configuration. 
 
 Following \cite{hodhigh}, it is convenient to introduce new 
 dimensionless variables 
 \begin{eqnarray}
 x=\frac{r-r_+}{r_+}\;,\;\;\tau=\frac{r_+-r_-}{r_+}\;,
 \end{eqnarray}
 in terms of which the radial equation (5) becomes 
 \begin{eqnarray}
 x(x+\tau)\frac{d^2 R}{dx^2}
 +(2x+\tau)\frac{dR}{dx}
 \nonumber\\+\left[q^2 Q^2 x(x+\tau)-l(l+1))\right]R=0\;,
 \end{eqnarray} 
 where we have submit the superradiance critical frequency $\omega=q\Phi_H$ 
 in the above equation. 
 
 This equation can be solved by Bessel function in the double limit 
 \begin{eqnarray}
 qQ\gg1\;,\;\;x\ll\tau\;.
 \end{eqnarray}
 In this asymptotic regime, the radial equation can be reduced to 
 \begin{eqnarray}
 x \frac{d^2 R}{dx^2} +\frac{dR}{dx}+ q^2 Q^2 x R=0\;. 
 \end{eqnarray} 
 The solution is then given by the Bessel function of the first 
 kind 
 \begin{eqnarray}
 R(x)=J_0(qQx)\;,
 \end{eqnarray} 
 i.e., the stationary scalar field is then described by the above function. 
 By taking account to the mirror-like boundary condition $R(x_m)=0$, we can obtain 
 the special mirror radius $r_m$ as 
 \begin{eqnarray}
 r_m=2M+\frac{j_{0,n}}{q\Phi_H}\;, \;\;\;n=1,2,3,\cdots
 \end{eqnarray}
 where $j_{0,n}$ is the $n$th positive zero of the Bessel function $J_0(x)$. 
 From this expression, we can see that, when $qQ\gg1$, 
 the reflecting mirror should be placed very near the horizon 
 to form the cloud configuration. This is consistent with the 
 near horizon condition $x\ll\tau$.

 \section{Conclusion}

 In summary, in this paper, we have studied
 the massless scalar clouds in the charged stringy black holes
 with the mirror-like boundary conditions.
 The scalar clouds are stationary bound states satisfying
 the superradiant critical frequency $\omega=q\Phi_H$.
 The scalar clouds in rotating black holes \cite{herdeiroprl,benone}
 can be heuristically interpreted
 in terms of a mechanical equilibrium between the
 Black hole-cloud gravitational attraction and angular momentum
 driven repulsion. For the charged black hole cases,
 the charged clouds can not be formed because gravitational attraction
 and electromagnetic repulsion can not reach equilibrium \cite{RNclouds}.
 Additional mirror should be placed at special location
 to reflect the charged scalar wave.

 We show that, for the specific set of
 black hole and scalar field parameters, the clouds are only possible for
 the specific mirror location $r_m$. For example,
 for the fixed parameters of black hole and scalar field $M, Q, q, l$,
 the discrete set of the mirror location $r_m$ is characterized by the
 node number $n$ of the radial wave function. It is shown that,
 the analytical results of mirror location $r_m$ for the clouds
 are perfectly coincide with the numerical results in the region of $qQ\ll 1$.
 However, the agreement becomes less impressive for $qQ=O(1)$ values.
 In addition, we also show that
 the massless scalar clouds are also possible when the mirror locations are
 very close to the horizon. At last, we present an analytical calculation
 of the specific mirror locations $r_m$ for the scalar clouds in the $qQ\gg 1$ regime.

 \section*{ACKNOWLEDGEMENT}

 The authors would like to thank Dr. Hongbao Zhang
 for useful discussion on the numerical methods.
 This work was supported by NSFC, China (Grant No. 11205048).


\begin{thebibliography}{99}


 \bibitem{hodprd2012}

 S. Hod, Phys. Rev. D \textbf{80}, 104026 (2012).


 \bibitem{hodepjc2013}

 S. Hod, Eur. Phys. J. C \textbf{73}, 2378 (2013).

 \bibitem{herdeiroprl}

 C. A. R. Herdeiro, and E. Radu, Phys. Rev. Lett. \textbf{112}, 221101 (2014).


 \bibitem{herdeiroprd}

 C. A. R. Herdeiro, and E. Radu, Phys. Rev. D \textbf{89}, 124018 (2014).
 
 \bibitem{1501}
 
 C. Herdeiro, and E. Radu, arXiv:1501.04319[gr-qr].
 
 

 \bibitem{hodprd2014}

 S. Hod, Phys. Rev. D \textbf{90}, 024051 (2014).

 \bibitem{RNclouds}

 J. Degollado, and C. Herdeiro, Gen. Rel. Grav. \textbf{45}, 2483 (2013).

 \bibitem{benone}

 C. Benone, L. Crispino, C. Herdeiro, and E. Radu,
 Phys. Rev. D \textbf{90}, 104024 (2014).

 \bibitem{sampaio}

 M. Sampaio, C. Herdeiro, and M. Wang,
 Phys. Rev. D \textbf{90}, 064004 (2014).

 \bibitem{grahamprd}

 A. Graham, and R. Jha, Phys. Rev. D \textbf{90}, 041501 (2014).

 \bibitem{Degolladoclouds}

 J. Degollado, and C. Herdeiro, Phys. Rev. D \textbf{90}, 065019 (2014).

 \bibitem{Brihaye}

 Y. Brihaye, C. Herdeiro, and Eugen Radu, Phys. Lett. B \textbf{739}, 1 (2014).

 \bibitem{HRRPLB}

 C. Herdeiro, E. Radu, and H. Runarsson, Phys. Lett. B \textbf{739}, 302 (2014).

 \bibitem{hodplb1}

 S. Hod, Phys. Lett. B \textbf{739}, 196 (2014).

 \bibitem{hodplb2}

 S. Hod, Phys. Lett. B \textbf{736}, 398 (2014).

 \bibitem{acoustic}

 C. Benone, L. Crispino, C. Herdeiro, and E. Radu, arxiv: 1412.7278 [gr-qc].

 \bibitem{bardeen}

 J. M. Bardeen, W. H. Press, and S. A. Teukolsky,
 Astrophys. J. \textbf{178}, 347 (1972).

 \bibitem{misner}

 C. W. Misner, Bull. Am. Phys. Soc. \textbf{17}, 472 (1972).

 \bibitem{bekenstein}

 J.D. Bekenstein,  Phys. Rev. D \textbf{7}, 949 (1973).

 \bibitem{liprd}

 R. Li, Phys. Rev. D \textbf{88}, 127901 (2013).


 \bibitem{detweiler}
 S. Detweiler, Phys. Rev. D \textbf{22}, 2323 (1980); H. Furuhashi
 and Y. Nambu, Prog. Theor. Phys. \textbf{112}, 983 (2004).


\bibitem{press}

 W. H. Press, and S. A. Teukolsky, Nature (London) \textbf{238},
 211 (1972).

 \bibitem{cardoso2004bomb}

 V. Cardoso, O. J. C. Dias, J. P. S. Lemos, and S. Yoshida,
 Phys. Rev. D \textbf{70}, 044039 (2004).


 \bibitem{liepjc2014}

 R. Li, and J. Zhao, Eur. Phys. J. C \textbf{74}, 3051(2014).

 \bibitem{liplb2015}

 R. Li, and J. Zhao, Phys. Lett. B \textbf{740}, 317 (2015).

 \bibitem{GM}

 G. W. Gibbons, and K. Maeda, Nucl. Phys. B \textbf{298}, 741 (1998).


 \bibitem{GHS}

 D. Garfinkle, G. T. Horowitz, and A. Strominger, Phys. Rev. D \textbf{43}, 3140 (1991).

 \bibitem{dilatonsr}
 K. Shiraishi, Mod. Phys. Lett. A \textbf{7}, 3449 (1992);
 J. Koga and K. Maeda, Phys. Lett. B \textbf{340}, 29 (1994).

 \bibitem{Degolladoprd}

 J. C. Degollado, C. A. R. Herdeiro, and H. F. Runarsson,
 Phys. Rev. D \textbf{88}, 063003 (2013).

 \bibitem{Dolanprd2010}

 S. R. Dolan, L. A. Oliveira, and L. C. B. Crispino, Phys. Rev. D \textbf{82}, 084037(2010).

 \bibitem{Cardosoprd2014}

 L. A. Oliveira, V. Cardoso, and L. C. B. Crispino, Phys. Rev. D \textbf{89}, 124008(2014).

 \bibitem{Uchikata}

 N. Uchikata, S. Yoshida, and T. Futamase, Phys. Rev. D \textbf{80}, 084020(2009).

 \bibitem{barranco}

 J. Barranco, et.al. Phys. Rev. Lett. \textbf{109}, 081102 (2012).
 
 \bibitem{hodhigh} 
 
  S. Hod, Phys. Rev. D \textbf{88}, 064055 (2013).


 \end{thebibliography}
 \end{document}